\documentclass[12pt]{spieman}  
\usepackage{amsmath,amsfonts,amssymb}
\usepackage{graphicx}
\usepackage{setspace}
\usepackage{tocloft}
\usepackage{hyperref}
\usepackage{algorithmic}
\usepackage{graphicx}
\usepackage{textcomp}
\usepackage{romannum}
\usepackage{booktabs, multirow, adjustbox}
\usepackage{lineno}
\usepackage{color,soul}
\usepackage{cite}

\title{Deep learning-based method for segmenting epithelial layer of  tubules in histopathological images of testicular tissue}

\author[a,*,\textdagger]{Azadeh Fakhrzadeh}
\author[b,\textdagger]{Pouya Karimian }
\author[b,\textdagger]{Mahsa Meyari }
\author[c]{Cris L. Luengo Hendriks}
\author[d]{Lena Holm}
\author[e]{Christian Sonne}
\author[e]{Rune Dietz}
\author[f]{Ellinor Spörndly-Nees}
\affil[a]{Information Technology Department, Iranian Research Institute for Information Science and Technology(IranDoc), Enghelab, Tehran, Iran, 1315773314}
\affil[b]{Industrial Engineering and Management Systems Department,Amirkabir University of Technology(Tehran Polytechnic), Tehran, Iran,1591634311 }
\affil[c]{Deepcell, Inc., 4025 Bohannon Dr., Menlo Park, California, USA, CA 94025}
\affil[d]{Department of Anatomy, Physiology, and Biochemistry, Swedish University of Agricultural Sciences, Uppsala, Sweden}
\affil[e]{Arctic Research Centre (ARC), Department of Ecoscience, Aarhus University, P.O. Box 358, DK-4000, Roskilde, Denmark}
\affil[f]{Department of pathology and wildlife Diseases, National Veterinary Institute(SVA), Uppsala, Sweden}

\cftpagenumbersoff{figure}
\cftpagenumbersoff{table} 

\begin{document}
 \maketitle

\begin{abstract}
\\
\textbf{Purpose:}There is growing concern that male reproduction is affected by environmental chemicals. One way to determine the adverse effect of environmental pollutants is to use wild animals as monitors and evaluate testicular toxicity using histopathology. Automated methods are necessary tools in the quantitative assessment of histopathology to overcome the subjectivity of manual evaluation and accelerate the process. We propose an automated method to process histology images of testicular tissue.\\
\textbf{Approach:} Testicular tissue consists of seminiferous tubules and interstitial tissue with the epithelium of seminiferous tubules containing cells that differentiate from primitive germ cells to spermatozoa in several steps.  Segmenting the epithelial layer of the seminiferous tubule is a prerequisite for developing automated methods to detect abnormalities in tissue. We suggest an encoder-decoder fully connected convolutional neural network (F-CNN)  model to segment the epithelial layer of the seminiferous tubules in histological images. The  ResNet-34  is used in the feature encoder module, which adds a shortcut mechanism to avoid the gradient vanishing and accelerate the network convergence. The squeeze \& excitation (SE) attention block is integrated into the encoding module improving the segmentation and localization of epithelium.\\
\textbf{Results:} We applied the proposed method for the 2-class problem where the epithelial layer of the tubule is the target class. The f-score and IoU of the proposed method are ${0.85\%}$ and ${0.92\%}$.  Although the proposed method is trained on a limited training set, it performs well on an independent dataset and outperforms other state-of-the-art methods. \\
\textbf{Conclusion:} The pretrained ResNet-34 in the encoder and attention block suggested in the decoder result in better segmentation and generalization. The proposed method can be applied to testicular tissue images from any mammalian species and can be used as the first part of a fully automated testicular tissue processing pipeline. The dataset and codes are publicly available on GitHub.
\end{abstract}
\keywords{Segmentation, Deep learning, Histological image,  Seminiferous tubules}

{\noindent \footnotesize\textbf{*}Azadeh Fakhrzadeh, \linkable{fakhrzadeh@irandoc.ac.ir}}\\
{\noindent \footnotesize\textbf{\textdagger} These authors contributed equally to this work}


%
%
%
%


\section{Introduction}
\label{sec:introduction}
Histopathology of testicular tissue has been considered the most sensitive tool to detect adverse effects from e.g. environmental chemicals on the male reproductive tract \cite{lanning2002recommended}. Increased frequency of reproductive disturbance is one of the health concerns and a large number of studies, reviewed by WHO, linked adverse effects on male reproduction in mammals to endocrine-disrupting environmental pollutants \cite{bergman2013state}. Testicular tissue consists of seminiferous tubules and interstitial tissue, segmenting the epithelium of tubules and classifying them into stages based on defined cell associations assist the pathologist in the interpretation of the sectioned tissue and detecting abnormalities. Manual analysis is subjective and may vary between different experts conducting the work. One crucial prerequisite step in the computerized analysis of histological images is the accurate detection of different objects of interest including specific cells, epithelium, and tubules. By segmenting the epithelium of seminiferous tubules, many automated quantification methods are applicable to single tubules to understand tissue abnormalities. The Society of Toxicological Pathology recommends classifying the testicular epithelium into stages when assessing tissue damage and determining imbalances in spermatogenesis \cite{pelletier1986cyclic}. Automated segmentation help to generate richer datasets for quantification and classification of tubules, assisting toxico-pathological examinations.

Preparation procedure, improper cutting or imaging, and different stages of development affect the shape and texture of seminiferous tubules in histological images. All of these variations make segmentation a very challenging task and traditional image segmentation methods based on handcrafted features are insufficient. Deep learning methods have achieved impressive results in computer vision tasks. Deep learning methods are based on neural networks and large annotated image datasets are needed to train them. One of the main challenges in employing deep learning methods for medical image analysis is limited high-quality training data with accurate annotations. Complicated acquisition and preparation procedure, and difficulties of manually annotating all the objects of interest in medical images are some of the obstacles to generating a sufficient training set.

 Here we suggest an encoder-decoder fully connected convolutional neural network (F-CNN)  model to segment the epithelial layer of the seminiferous tubules in histological images. The  ResNet-34 \cite{he2016deep} is used in the feature encoder module, which adds shortcut mechanism to avoid the gradient vanishing and accelerate the network convergence. We integrate the attention block suggested in Ref.~\citenum{roy2018recalibrating} within F-CNNs, in the encoding module. Hu et al.\cite{hu2018squeeze} introduced squeeze \& excitation (SE) attention block which factors out the spatial dependency by global average pooling.  Their suggested method learns a channel specific descriptor, which is used to rescale the input feature map to highlight only useful channels. The CNN model with such SE blocks achieved the best performance in the ILSVRC 2017 image classification competition on the ImageNet dataset, indicating its efficiency \cite{hu2018squeeze}. By applying global average pooling every intermediate layer has the total receptive field of the input image. The pixel-wise spatial information is more informative for fine-grained segmentation tasks of highly complex structures, such as medical images. Roy et al. \cite{roy2018recalibrating} introduced an alternate SE block, which "squeezes" along the channels and "excites" spatially. This is complementary to the SE block, as it does not change the receptive field, but provides spatial attention to focus on certain regions. For the segmentation, they suggest an attention block which is a combination of these spatial and channel "squeeze \& excitation"  blocks within a single block to recalibrate the feature maps separately along channel and space. The main contribution of this paper can be summarized
as follows:
\begin{itemize} 
\item An automated method to segment the epithelial layer of seminiferous tubules is suggested. 
\item The suggested network is trained on testicular tissue from Mink stained with periodic acid-Shiff (PAS) and validated on testicular tissue from polar bear stained with hematoxylin and eosin (H\&E). Our algorithm obtains highly satisfactory results and outperforms state-of-the-art methods.
\item The pretrained ResNet-34 in the encoder and attention block suggested in the decoder result in better segmentation and generalization. The suggested can be applied to histopathological images of testicular tissue from any mammalian species with any choice of staining.
\item The suggested method can be used as the first part of a fully automated testicular tissue processing pipeline.
\item The training set (1072 manually segmented tubules) and the codes and weights of networks  are publicly available.  

\end{itemize} 

\section{Related works}
In traditional image processing, handcrafted features based on intensity of pixels and spatial relationship between them are derived from the image. These features are used to detect regions of interest such as cells, glands, tubular structure and epithelial tissues and to classify them. Sirinukunwattana el al. \cite{sirinukunwattana2015novel} used texture and color of  each superpixel to estimate the probability of superpixels belonging to glandular regions, resulting in a glandular probability map. Niak et al. \cite{naik2007gland} used Bayesian classifier to detect candidate gland regions by utilizing low-level image features, while Altunbay et al. \cite{altunbay2009color} constructed a graph on multiple tissue components and colored its edges depending on the component types of their endpoints. Subsequently, a new set of structural features re-derived from color graphs and used this in the classification of tissues. Other methods take advantage of prior shape information of histological structures and applied mathematical models to detect objects of interest. For example, Fakhrzadeh et al. \cite{fakhrzadeh2012analyzing} employed geodesic distance transform to detect tubular boundaries, and Fu et al. \cite{fu2014novel} used polar space random field model to segment glandular structures. 

Advances in deep learning have resulted in new methods  which generally outperform handcrafted features on segmentation tasks in digital pathology  \cite{bardou2018classification, xu2016deep}. Xu et al. \cite{xu2019histopathological} applied deep residual network (ResNet) on patches extracted from Whole slide imaging (WSI) of testicular tissue section to segment seminiferous tubules.  Cire\c{s}san et al. \cite{cirecsan2013mitosis} trained a supervised deep neural network to differentiate patches with a mitotic nucleus from all other patches. To detect cancer metastases in breast atypical lymph nodes at a fine-grained level, Wang et al. \cite{wang2016deep} applied convolutional neural network (CNN) on millions of small positive and negative patches from WSI to assign a prediction score to every patch. The final decision is aggregated from the micro predictions. BenTaieb et al. \cite{bentaieb2016multi} proposed a multi-loss convolution network that performs both classification and segmentation of adenocarcinoma glands using a joint learning of a segmentation and classification that is modelled in a unified framework based on a novel deep learning architecture and multi-loss objective function. Chen et al. \cite{chen2016dcan} proposed contour-aware deep learning architecture network (DCAN) under a unified multi-task learning framework for more accurate detection and segmentation of glandular structure. Xu et al. \cite{xu2016deep} designed a deep multichannel side supervision system (DMCS), where region and boundary cues were fused with side supervision to segment glands in colon histological images. Xu et al. \cite{xu2019histopathological} used deep residual network (ResNet) to segment seminiferous tubule from mouse (\textit{Mus musculus}) testicular sections. Then, utilizing other deep ResNet for multi-cell; spermatid, spermatocyte, and spermatogonia, segmentation and a fully convolutional network (FCN) for multi-region; elongated spermatid, round spermatid, spermatogonial and spermatocyte region segmentation. Zhang et al. {\cite{zhang2022discriminative}} suggested a label rectification method based on error correction, namely ECLR, which can be directly added after the fully-supervised segmentation framework. They proposed a collaborative multi-task discriminative error prediction network (DEP-Net) and the specific mask degradation methods to high-light the inter-class error and intra-class error, enhancing the error recognition ability of DEP-Net. Their suggested method performed well on gland segmentation.

\begin{figure*}[t]
\centerline{\includegraphics[width=\textwidth]{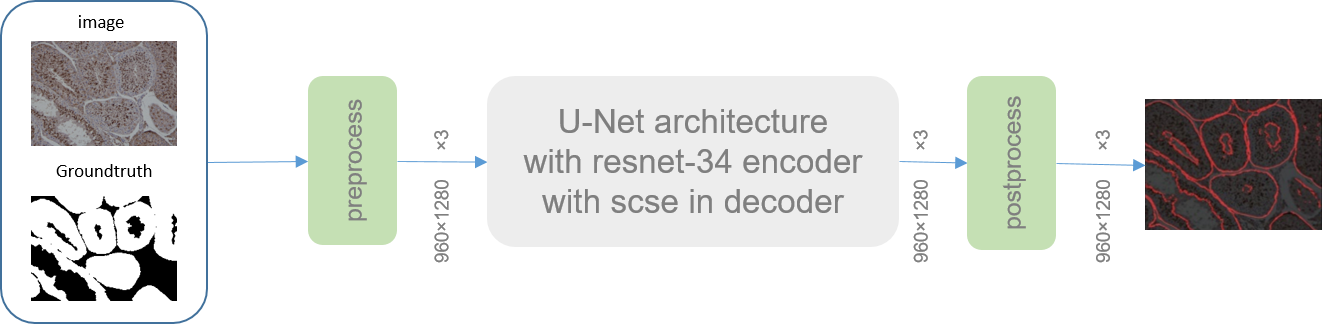}}
\caption{Overview of proposed method. The inputs  are RGB images with the size of $960 \times 1280$ and their corresponding ground truth. In Preprocessing  data are  augmented and normalized and then  encoder-decoder network is applied. In postprocessing, seeded watershed algorithm is used to separate touching objects.}
\label{fig:method}
\end{figure*}

Recently, most of the state of the arts CNNs for segmentation consist of two parts: encoder, which extracts feature maps, and decoder, which produces segmentation map by upsampling feature maps. One of the first networks introduced based on encoder-decoder architecture was U-Net \cite{ronneberger2015u}. The encoder section in U-Net takes the input image performing operations such as convolutions and pooling to capture rich contextual information at different resolution levels. The decoder section uses features obtained from encoder section to localize objects of interest and obtains sharp object boundaries. U-Net proved to be effective in many medical image segmentation tasks, specifically when available data is limited.
 Salvi et al. {\cite{salvi2021hybrid}} designed a glands segmentation strategy using a multi-channel algorithm that exploits and fuses both traditional and U-Net based deep learning techniques. Specifically, their proposed approach employs a hybrid segmentation strategy based on stroma detection to accurately detect and delineate the prostate glands contours.  
 Bouteldja et al. {\cite{bouteldja2021deep}} proposed an automated algorithm based on U-Net to accurate segmentation of periodic acid-Schiff-stained kidney tissue. Salvi et al. {\cite{salvi2021automated}} proposed a strategy combining the accuracy of a level-set with the semantic segmentation of U-Net to detect the glomeruli and tubules contours in histopathological images.

Many new medical image segmentation networks used U-Net as their base architecture and tried to improve it. Zhou et al. {\cite{zhou2018unet++}} introduced U-Net++ and extended skip connections by adding more convolutions into them. Applying a combination of recurrent and residual blocks into a U-Net architecture resulted in R2U-Net, which can benefit from the advantages of residual and recurrent blocks{\cite{alom2019recurrent}}. Although both methods outperform U-Net, they are more complicated and more challenging to train when data is limited. Roy et al.\cite{roy2018recalibrating} improved U-Net by applying the attention concept in the decoder part. In PSPNet \cite{zhao2017pyramid} and Deeplabv3+ \cite{chen2018encoder} first feature maps are generated by a known CNN such as ResNet \cite{he2016deep}. In PSPNet \cite{zhao2017pyramid}, a pyramid pooling module suggested to make hierarchical layers of information by using different sizes of convolution. This module combines feature maps under different pyramid scales. The outputs of convolution and upsampling layers concatenated with initial feature maps to generate segmentation map. Pyramid pooling module creates feature maps with diverse resolutions. Deeplabv3+ \cite{chen2018encoder} is based on PSPNet being at the end of encoder Atrous Spatial Pyramid Pooling (ASPP)  is employed. ASPP is a pyramid pooling module using atrous convolution \cite{chen2017rethinking} opposite of simple convolution. Atrous convolution is a useful tool for semantic segmentation because of effective upsampling. Although both PSPNet and Deeplabv3+ work well for semantic segmentation, limited data for training could be a challenge for them.

Histopathology is used to detect adverse effects on male reproduction . During spermatogenesis, stem cells (spermatogonia) constantly divide in a dynamic organised process, resulting in motile haploid spermatozoa. It is known that various steps in spermatogenesis are affected by environmental pollutants {\cite{saradha2006effect}}, but the complexity of testicular tissue makes histopathological evaluation difficult and time-consuming {\cite{creasy2001pathogenesis, berndtson2011importance}}. Image analysis has been discussed recently as a future necessary tool in quantitative evaluation of histopathology to overcome the subjectivity of manual evaluation and accelerate the process {\cite{nature2012quest}}. When assessing tissue damage, seminiferous epithelium needs to be classified into different stages to detect certain cell damages; but stage identification is a demanding task. Segmentation of the tubular epithelium  is an essential first step in automated evaluation and staging of tubules and to the best of our knowledge no segmentation method has been suggested for this purpose. Here we suggest a novel method to segment tubular epithelium of testicular tissue in various animal species.

\section{Dataset}

Training set are images of testis from five, sexually mature mink were collected at the annual culling on a mink farm. No ethical approval was required because of the use of routinely culled mink from a fur farm. Testis were fixed in 4\% formalin and embedded in paraffin. Immunohistochemical localization of Gata-4  was identified in sections from all five mink by E Sporndly-Nees, DVM, trained in histological evaluation of testicular tissue described in details elsewhere \cite{sporndly2015effect}. The sections were stained with a variant of periodic acid-Shiff (PAS) staining to obtain maximum colour differences between the seminiferous epithelium and the interstitial tissue. Digital images of the sections were taken using a Nikon Eclipse 50i microscope and Nikon Digital Sight DS-2M camera, using the 20 objective lens. Images were stored as uncompressed TIFF files, at  $1200\times 1600$ pixels and $0.4$ mm per pixel, as red–green–blue images with 8 bits per channel. 

The test data originate from  six polar bears (\textit{Ursus maritimus}) testis. The samples were collected from the aboriginal subsistence hunting in Scoresby Sound, East Greenland during the period 1997-2016, regulated by annual quotas by Greenland Home Rule,  Nuuk since 2006. The testis were fixed in formaldehyde/alcohol solution and processed for conventional histology, embedded in paraffin, sectioned (4 um) and stained with hematoxylin and eosin (H\&E). Detailed information of the polar bears can be seen in our previous publication \cite{sporndly2019age}. Digital images were taken and stored as described above.

We used one section and one slide per animal, the images were taken in a grid likes structure to assure that no tubules were imaged twice.

\section{Method}
We suggest a segmentation method based on CNN. The structure of proposed method is similar to U-Net, and has two parts: encoder and decoder. First, the training set is generated and augmented and then suggested model is applied as shown in Fig. \ref{fig:method}.

\subsection{Training  and testing sets}
To delineate the borders of the epithelium we used the livewire algorithm implemented in Matlab \cite{chodorowski2005color}. The program determines various measures for each pixel belonging to an edge. From the gradient image, the gradient magnitude and gradient direction is calculated. The user added a seed point on the boundary of epithelium  of a tubule and the algorithm calculated the cost of the optimal boundary between this seed point and all other pixels in the image. By moving the mouse over the image, the user instantly saw the optimal boundary between that seed point and the mouse position. By clicking, that portion of the boundary were fixed, and the newly selected pixel became the seed point, repeating the whole process.
 The seeds are points on the boundaries of the tubules and their lumens. The boundaries are easy to recognize by a human observer. Different users would select the same boundaries. Manual segmentations was done by Mahsa Meyari and validated by Azadeh Fakhrzadeh. For two classes training set the epithelial layers are true positive. In three classes training set, class 1 is the epithelial layer, tubules borders are the class 2 and the rest of the image is the class 3. One hundred and sixteen images from mink testis stained with PAS was considered and 1072 tubules were manually segmented for training the network. Ten images from polar bear stained with H\&E was used and 93 tubules  were manually segmented for validating the network.

\subsection{Preprocessing}
\begin{figure}[t]
\centering
\includegraphics[width=\textwidth]{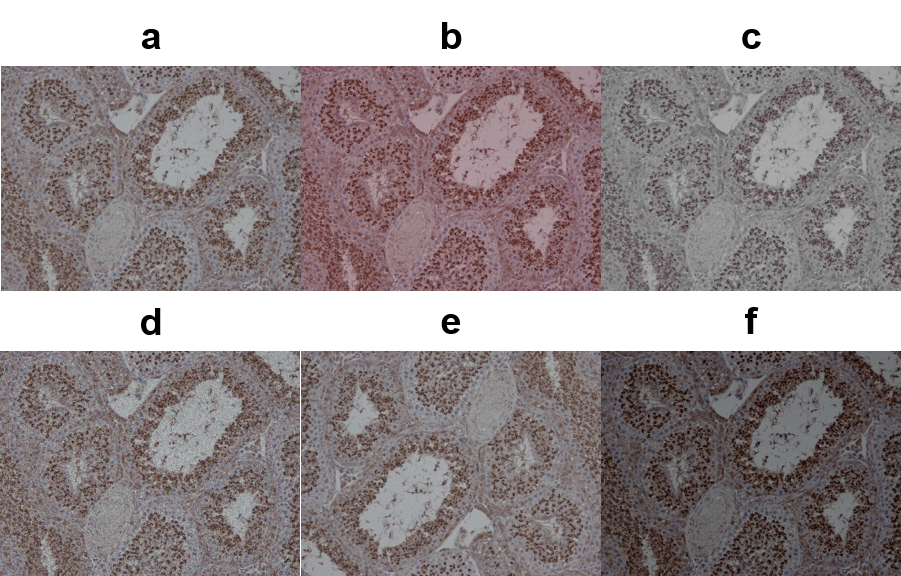}
\caption{Data augmentation transforms. (a) Original image (b) RGB shift (c) Changing HSV (d) Adding Gaussian noise (e) Horizontal and vertical flip (f) Changing brightness and contrast.}
\label{fig:augmentation}
\end{figure}

Medical image analysis methods often suffer from a shortage of data and this work is not an exception. Lack of sufficient data in training neural networks may cause statistically overfitting which prevents models from being trained for long epochs. Augmentation methods can partly  solve this problem. Augmentation methods that are applied comprise horizontal and vertical flip, adding Gaussian noise, random change of brightness, contrast, RGB random shift, and HSV random shift. Each transform applied with probability of 0.5 as shown in Fig. \ref{fig:augmentation}. 

After augmenting data, inputs are normalized using Imagenet pretrained weights, mean and standard deviation \cite{deng2009imagenet}. For training from scratch, mean and standard deviation of dataset is used.  Images are passed to network after data augmentation and normalization.

\subsection{Network and losses}
\begin{figure}[t]
\centering
\includegraphics[width=\textwidth]{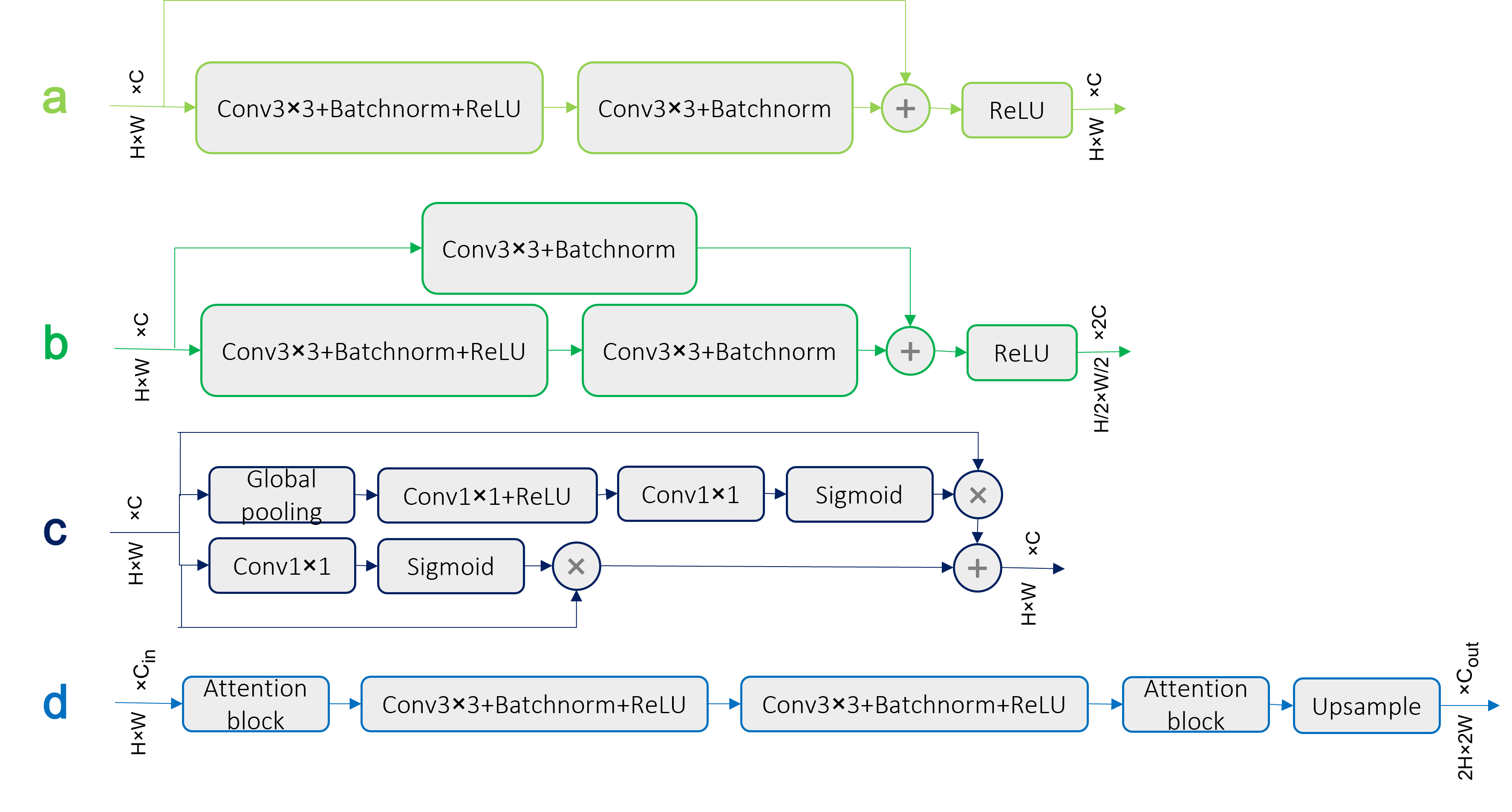}
\caption{Different blocks used in network architecture: (a) residual block (b) bottleneck block (c) attention block (d) decoder block}
\label{fig:blocks}
\end{figure}

Suggested method is based on U-Net \cite{ronneberger2015u}. U-Net is a U shape method that has two main parts; encoder and decoder. In encoder part feature maps are extracted  from the image and in decoder part feature maps will be upsampled to the original size of the image for dense prediction. The proposed method is different from U-Net in both encoder and decoder. Different convolutional blocks  are used in order to help in better feature extraction and generalization. All of the blocks of the network are illustrated in Fig. \ref{fig:blocks} and the following details of the network and loss function is  explained.

{\bf Encoder.} As for encoder  ResNet-34 \cite{he2016deep} is used. ResNet is a well-known backbone network that can extract features effectively. In computer vision usually networks with deeper layers are used. However, most of the deep architectures suffer from vanishing gradient in the first layers. Residual blocks in ResNet resolve this problem by adding activation from previous layers to the activation of deeper layers. ResNet-34 has two main blocks: residual block and bottleneck block. Details of these blocks are shown in Fig. \ref{fig:blocks}. The encoder part consists of five downsampling layers which are layers of ResNet-34.

\begin{figure}[!ht]
\centering
\includegraphics[width=\textwidth]{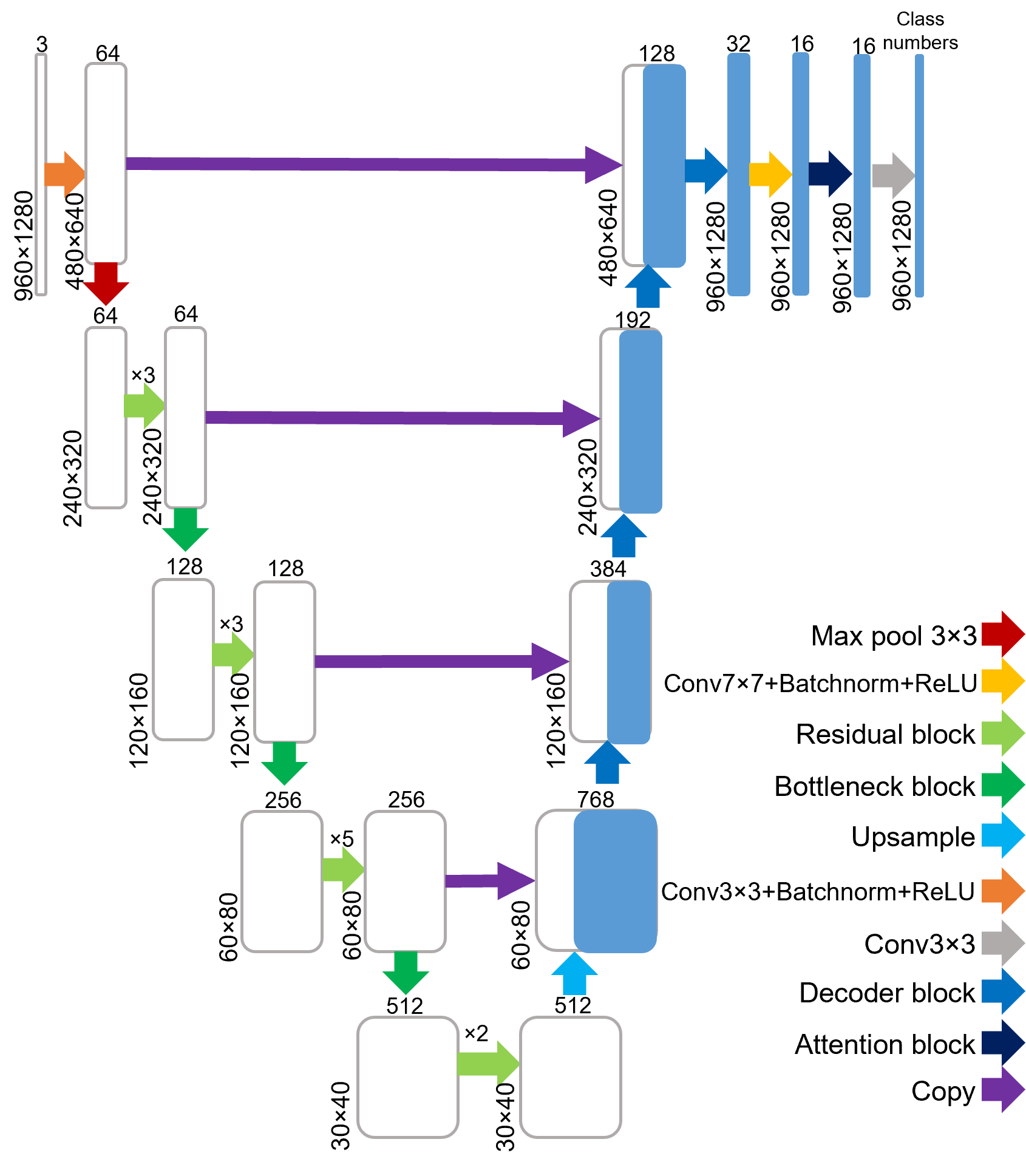}
\caption{Network architecture($\times n$ on top of a layer means that layer repeated n times)}
\label{fig:network}
\end{figure}

{\bf Decoder.} The decoder part upsamples the features extracted by the encoder. As shown in Fig. \ref{fig:network} the suggested decoder module consists of attention block, Conv$3\times3$, batchnorm \cite{ioffe2015batch}, ReLU activation, and upsampling layer. Instead of upconvolution in U-Net, we suggest nearest neighbor interpolation, which is less computationally expensive. In the attention  Spatial and Channel Squeeze \& Excitation Block(scSE) \cite{roy2018recalibrating} is used. The scSE is a combination of Spatial Squeeze \& Channel Excitation Block (cSE) and Channel Squeeze \& Spatial Excitation Block (sSE). The cSE block squeezes spatial information and the channels are recalibrated. The cSE contains global pooling, Conv$1\times1$, ReLU, Conv$1\times1$, and sigmoid to squeeze spatial information in channels. This squeezed information magnifies the best channels. The input of the block is refined by channel-wise multiplication of it with squeezed spatial information. The sSE block, squeezes channels and the receptive field is unchanged, applying Conv$1\times1$ and sigmoid to squeeze channels information. The input of the block is multiplied by squeezed channels information. The sSE block determines the most important regions of the image. The output of the attention block is obtained by element-wise adding of cSE and sSE. The attention block can demonstrate important sections of images and promotes essential features for better segmentation and generalization \cite{roy2018recalibrating}. The attention block and decoder block are illustrated in Fig. \ref{fig:blocks} parts c and d. At the end of the decoder, the following layers are used for producing the segmentation map: Conv$3\time3$, batchnorm, ReLU activation, attention block, and Conv$3\times3$. Batch normalization is used to improve the training process. The segmentation map could have two or three channels (classes). Network architecture is illustrated in Fig. \ref{fig:network}.

{\bf Loss Functions.} Cross Entropy loss is generally the default loss for a segmentation model. Another common loss function in segmentation tasks is Dice loss \cite{milletari2016v} which tries to optimize the Dice score. Combination of these two losses can have advantages of both losses \cite{taghanaki2019combo}. We used the  combination of Cross Entropy and Dice loss function as below:

\begin{equation}
I_{Dice + WCE} = l_{ce} + l_{dc}
\end{equation}

\begin{equation}
l_{ce} =  -\frac{1}{N} \sum_{n=1}^N  \sum_{c=1}^C w_c  y_{n,c}  \log \widehat{y}_{n,c}   
\end{equation}

\begin{equation}
l_{dc} = - \frac{2}{\lvert C \rvert} \sum_{c=1}^C  \frac{ \sum_n \widehat{y}_{n,c} y_{n,c}   } {\sum_n \widehat{y}_{n,c} + \sum_n y_{n,c}}
\end{equation}

Where $\widehat{y}_{n,c}$ is the softmax output of the network and $y_{n,c}$ is a one hot encoding of the groundtruth segmentation map. N and C are total number of samples and classes respectively. The related weight of every class $w_c$ is defined as $w_c = \frac{\sum_c x_c}{C \ x_c} $ that $x_c$ is the total number of class $c$ pixels.

One of the main challenges in medical images tasks is imbalanced datasets. Salehi et al. \cite{salehi2017tversky} suggested  Tversky loss function for imbalanced medical datasets. This loss function penalizes false prediction by using alpha and beta parameters. Alpha and beta parameters control the magnitude of false positives and false negatives respectively. Tversky loss function is defined as below:

\begin{align}
l_{tv} = \frac{-2}{\lvert C \rvert} \sum_{c=1}^C  \frac{ \sum_n \widehat{y}_{n,c} y_{n,c}   } { \sum_n \widehat{y}_{n,c} y_{n,c} +
\alpha \left(1-y_{n,c}\right)\widehat{y}_{n,c} + \beta \left(1-\widehat{y}_{n,c}\right)  y_{n,c}}
\end{align}

where $\alpha=0.3$ and $\beta=0.7$ based on the best performance metrics \cite{salehi2017tversky}. Both of these loss functions were tested. 

\subsection{Post processing}
Due to the improper cutting angle or poor fixation, some of tubules collapse masking boundary between those tubules. As a result, we may have touching objects in  the output of the network. To overcome this problem the seeded watershed method can be used. In this method, the input image is seen as a topographic surface and stimulates flooding from seed points. The seed points work as local minima of the image. Watershed builds a border between the seed points, where the sum of the intensity values at the border is maximum. We can choose seed points manually at the center of every touching tubule or in a place that results in better separation.

\section{Results}

\begin{figure*}[t]
\includegraphics[width=\textwidth]{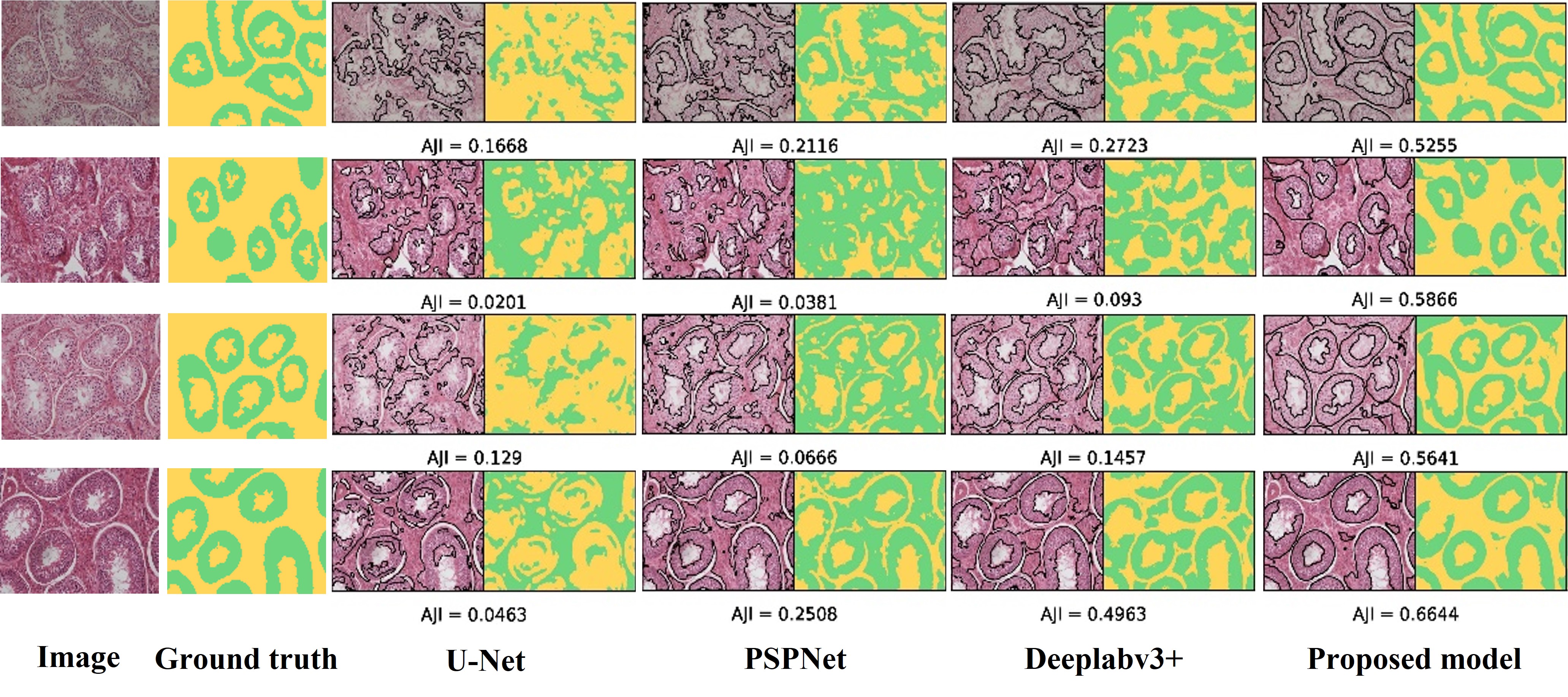}
\caption{Qualitative results of 2-class models on test data. Test data is  images from polar bear with H\&E staining. All models in the picture were trained on Mink images with PAS staining and  their loss function is  Tversky.}
\label{fig:staining}
\end{figure*}

Cross-validation decouples possible correlation between training and test sets by training multiple models on different subsets of the data, and testing each with data not seen by it during training. K-fold cross validation method with five folds is used to split data into five parts by random splitting to provide maximum use. Four parts are used for training and one part is used for validation in each fold. To evaluate models, mean and confidence interval at 95\% is calculated. Networks are trained for 60 epochs with the Adam optimizer \cite{kingma2014adam} and a learning rate of $10^{-4}$. The learning rate  decayed by a factor of 0.5 at epoch 30. In U-Net and ResNet-34, the input size should be divisible by $2^4$ and $2^5$  respectively, enabling the decoder to make an output with the same size as the input. As a result, encoder inputs in networks with ResNet-34 are resized to $960\times 1280$. In other networks, images are passed with the original size $1200\times 1600$. Also in ResNet-34 ImageNet pretrained weights are used. For training, we used a Quadro RTX 8000. The training time of the proposed network for 60 epochs is approximately 1 hour (noted that test time augmentation in validation is used and train batch size and validation batch size are respectively 4 and 2).  2-class outputs (epithelial layer and background) and 3-class outputs (epithelial layer, tubules boundary pixels and  background) are predicted. The purpose of considering 3-class is to differentiate tubules with touching borders as suggested in Ref.~\citenum{guerrero2018multiclass}. Although we applied cross validation, still the whole data set is similar. To evaluate the generality of methods we tested them on images from polar bear with different staining than training set. The evaluation metrics are  Intersection over Union (IoU), F-score and Mean AJI (Aggregated Jaccard Index). AJI is a metric designed for considering object-level and pixel-level errors at once {\cite{kumar2017dataset}}. It computes an aggregated intersection cardinality numerator, and an aggregated union cardinality denominator for all ground truth and segmented nuclei under consideration.

\textbf{Data Augmentation. }We applied data augmentation and test time augmentation. To better understand the augmentation effect, we applied three different augmentation options on a U-Net with 2 classes. These options are no augmentation, low data augmentation plus test time augmentation, and high augmentation plus test time augmentation. Data augmentation transforms are described in the method section. Implementation of data augmentation is done with Albumentations \cite{buslaev2020albumentations}. For low augmentation parameters of transformations are: Gauss noise mean=0, Gauss noise variance range=(0.4,0.6), RGB shift value=5, hue shift value range=(-2,2), saturation shift range=(-3,3), value shift range=(-2,2). In high augmentation, we increased the parameters of the transforms. Parameters for high augmentation are: Gauss noise mean=0, Gauss noise variance=1, RGB shift value=15, hue shift value range=(-20,20), saturation shift range=(-30,30), value shift range=(-20,20).Test time augmentation transforms consist of horizontal flip and vertical flip. Performance results of these three models  shown in Table \ref{table: augmentation}. Low augmentation model has the best results. Therefore, low augmentation used in all of models.

\begin{table}[b]
\centering
\caption{Comparing different level of augmentation}
\vspace{0.05in}

\label{table: augmentation}
\begin{tabular}{lcc}
\toprule
Model & Jaccard(IoU) & F-score \\ 
\midrule
U-Net(2-class) w/o augmentation & $0.85\pm0.02$ & $0.92\pm0.01$ \\ 
U-Net(2-class) w/ low augmentation & $0.86\pm0.03$ & $0.92\pm0.01$ \\ 
U-Net(2-class) w/ high augmentation & $0.71\pm0.02$ & $0.83\pm0.01$ \\ 
\bottomrule
\end{tabular}
\end{table}

\begin{table}[h]
\centering
\caption{Quantitative results of k-fold cross validation, for  mink tissue stained with PAS}
\label{table:results}
\vspace{0.05in}

\begin{adjustbox}{max width=\textwidth}
\begin{tabular}{llcccc}
\toprule
Class type & Model & Input & Loss & IoU & F-score  \\

\midrule
\multirow{8}{*}{2-class} & R2U-Net\cite{alom2019recurrent}&$1200\times 1600$&Tversky&$0.80\pm0.01$&$0.89\pm0.01$\\
                          &DeepLabv3+\cite{chen2018encoder}&$960\times 1280$&Tversky& $0.81\pm0.02$&$0.90\pm0.01$\\
						 &PSPNet\cite{zhao2017pyramid}&$960\times 1280$&Tversky&$0.82\pm0.02$&$0.90\pm0.01$\\
						 &U-Net\cite{ronneberger2015u}&$1200\times 1600$&Dice+WCE&$0.85\pm0.02$&$0.92\pm0.01$\\
						 &U-Net\cite{ronneberger2015u}&$1200\times 1600$&Tversky&$0.86\pm0.03$&$0.92\pm0.01$\\
						 &Proposed model&$960\times 1280$&Dice+WCE&$0.85\pm0.02$&$0.92\pm0.01$\\
						 &Proposed model w/o attention&$960\times 1280$&Tversky&$0.84\pm0.02$&$0.91\pm0.01$\\
						 &Proposed model&$960\times 1280$&Tversky&$0.84\pm0.02$&$0.91\pm0.01$\\
\hline
\multirow{4}{*}{3-class} &U-Net\cite{ronneberger2015u}&$1200\times 1600$&Dice+WCE&$0.78\pm0.02$&$0.87\pm0.01$\\
						 &U-Net\cite{ronneberger2015u}&$1200\times 1600$&Tversky&$0.82\pm0.02$&$0.90\pm0.01$\\
						 &Proposed model&$960\times 1280$&Dice+WCE&$0.76\pm0.03$&$0.87\pm0.01$\\
						 &Proposed model&$960\times 1280$&Tversky&$0.78\pm0.02$&$0.88\pm0.01$\\
\bottomrule
\end{tabular}
\end{adjustbox}
\end{table}

\begin{table}[h]
\centering
\caption{Quantitative results of 2-class, for polar bear tissue stained with H\& E}
\label{table: generalization}
\vspace{0.05in}
\begin{adjustbox}{max width=\textwidth}
\begin{tabular}{lccccc}

\toprule
Model & Input & Loss & IoU & F-score & Mean AJI\\

\midrule
R2U-Net\cite{alom2019recurrent}&$1200\times 1600$&Dice+WCE&$0.39$&$0.56$&$0.07$\\
U-Net\cite{ronneberger2015u}&$1200\times 1600$&Dice+WCE&$0.39$&$0.56$&$0.08$\\
U-Net\cite{ronneberger2015u}&$1200\times 1600$&Tversky&$0.39$&$0.56$&$0.08$\\
PSPNet\cite{zhao2017pyramid}&$960\times 1280$&Tversky&$0.46$&$0.63$&$0.10$\\
DeepLabv3+\cite{chen2018encoder}&$960\times 1280$&Tversky&$0.48$&$0.65$&$0.22$\\
Proposed model&$960\times 1280$&Dice+WCE&$0.65$&$0.78$&$0.45$\\
Proposed model w/o attention&$960\times 1280$&Tversky&$0.63$&$0.75$&$0.46$\\
Proposed model&$960\times 1280$&Tversky&$0.65$&$0.79$&$0.52$\\

\bottomrule
\end{tabular}
\end{adjustbox}
\end{table}

\section{Discussion}
As we can see in table \ref{table:results}, for 2-class U-Net, the results of both loss functions are close, with Tversky loss function works slightly better. In 3-class U-Net Tversky loss function gives better results. The best results of U-Net are for  2-class with Tversky  loss function, where mean of IoU and F-score for five folds are $0.86$ and $0.92$ respectively. For 2-class proposed model, Dice and weighted cross entropy loss function works better. In 3-class proposed model, Tversky loss function gives a better result. In both U-Net and the proposed method 2-class gives better qualification results, hence adding the boundary of tubules as a third class is not useful. To understand the effect of attention block we have tested the proposed method without attention block.  The best results of the proposed method are for 2-class with Dice and weighted cross entropy loss function where IoU and F-score are $0.85$ and  $0.92$ respectively. For both DeepLabv3+ \cite{chen2018encoder} and PSPNet \cite{zhao2017pyramid}, ResNet-34 is used as the encoder. It seemed that the shortage of data causes complicated models with high number of parameters, such as  DeepLabv3+ PSPNet and R2U-Net models, to produce worse results than U-Net and the proposed model.

Any mammalian testis is composed of the same type of tubules with the same structure in histological images. There is more variation among tubules within a single section from one testis, than there is variation among animals. Variation comes from the variation in the angle that each tubule is sectioned at, and the variation in curvature of the tubule at the point it is sectioned. This is the variation that we need to capture in our tubule segmentation model. Using cross validation, it is shown in table \ref{table:results} that our model is independent of the morphology of tubules. Another factor affecting appearance of tissues is type of staining. To show that our method is not dependent to the type of animal and type of staining, we applied  our trained model on polar bear tissues with H \& E staining and results are reported in  table \ref{table: generalization} and figure \ref{fig:staining}.  All the models in table \ref{table: generalization} and figure \ref{fig:staining} were trained on mink tissue with PAS staining. We see in table \ref{table: generalization} that, the proposed method with Tversky loss function, with IoU, F-score and Mean AJI of  $0.65$, $0.79$ and $0.52$ respectively,  performs better than other methods on test data.  Mean AJI shows how well the segmented objects are separated.  Also we can see that  proposed model with attention block with AJI $0.52$ works better than proposed model without attention block with AIJ $0.46$. This shows  that attention block results in better generalization. The qualitative results  of all the models  are shown in figure \ref{fig:staining}. Every row in this figure is a sample image from a different polar bear.  The color of every sample in figure \ref{fig:staining} looks different but in all of them the suggested model trained on PAS staining can detect the epithelium layer. Other models have poor performance.  These results suggest that the proposed method can work on histological images of testicular tissue of any species with different type of staining from the training set. Although this type of investigations is not generally done on human tissue, we do not see any limitation to apply our method to human tissue.

\section{conclusion}
Quantitative analysis of histological images of testicular tissue is an important tool in toxicology. Testicular tissue is composed of interstitial tissue and seminiferous tubules. In the tubules the germ cells are developing to become elongated spermatids. Each tubule is in a specific stage, with a unique composition of various combinations of developing germ cells, and each stage has a unique appearance due to toxic damage. To detect stage of every tubule automatically in order to quantitatively study the structure of tubules we first need to segment them. We have suggested a novel automated method based on a fully connected convolutional neural network for segmenting epithelium of tubules in histological images of testicular tissue. The proposed method has comparable metrics with U-Net, using cross-validation. We also tested the proposed method on independent test data with different staining than the training dataset and unlike U-Net, it performed well in detecting tubules.  Additionally, in comparison to U-Net, the suggested model has fewer parameters and is faster in inference. Because of its good generalization property, the proposed method can work just as well on testicular tissue from any mammalian species and likely avian as well, due to structural similarities, and with any choice of staining.  Using the suggested segmentation method, automated quantification analysis of the testicular tissue on a large scale is possible, leading to more reliable research outcomes. One of the limitations of our model is magnification of the microscope.  In order to use our trained model on a new dataset, the magnification of the microscope for the new dataset, needs to be the same as the data used here. Our method may fail to perform on tissue with disturbances to the extent that tubules do not have a consistent structure. To get better results with the proposed model, we suggest adding a few images of the new dataset to the training set and training the model again. Number of classes and type of loss function depends on the dataset.

\subsection*{Acknowledgment}
We are grateful for excellent technical assistance for histological preparations of Gunilla Ericson-Forslund and Astrid Gumucio. 
 Danish Cooperation for Environment in the Arctic (DANCEA), The Commission for Scientific Research in Greenland (KVUG), The Prince Albert II Foundation and the Arctic Research Centre (ARC) at Aarhus University are acknowledged for financial support. The study was also part of the International Polar Year (IPY) BearHealth project (IPY 2007–2008, Activity \#134) funded by the Independent Research Fund Denmark. Furthermore, we acknowledge the subsistence hunters in Scoresby Sound for obtaining samples.  
No ethical approval was required.We have no conflicts of interest to disclose.\\

\subsection*{Data, Materials, and Code Availability}
The archived version of the code and the generated data can be freely accessed through a Github repository: \linkable{\url{https://github.com/pouyaka/tube_segmentation}}.


\bibliography{epitheliumsegvrevisedfinal} 
\bibliographystyle{spiejour}   

\end{document}